\documentclass[english,prl,twocolumn,superscriptaddress,showpacs,letterpaper]{revtex4}

\usepackage{ae} 
\usepackage[T1]{fontenc}
\usepackage[ansinew]{inputenc}
\usepackage{amsmath}
\usepackage{amssymb}
\usepackage{graphicx}
\usepackage{color}
\usepackage{epsfig}
\usepackage{booktabs}

\newcommand{\Rmp}{R^{-(+)}}
\newcommand{\Rpp}{R^{+(+)}}
\newcommand{\Rmm}{R^{-(-)}}
\newcommand{\Rpm}{R^{+(-)}}
\newcommand{\Bmp}{B^{-(+)}}
\newcommand{\Bpp}{B^{+(+)}}

\newcommand{\Bpm}{B^{+(-)}}

\newcommand{\beq}{\begin{equation}}
\newcommand{\eeq}{\end{equation}}
\newcommand\beqa{\begin{eqnarray}}
\newcommand\eeqa{\end{eqnarray}}
\newcommand\bea{\begin{array}}
\newcommand\eea{\end{array}}

\def\XXint#1#2#3{{\setbox0=\hbox{$#1{#2#3}{\int}$}
\vcenter{\hbox{$#2#3$}}\kern-.5\wd0}}

\newcommand{\nn}{\nonumber}

\newcommand{\COMMENT}[1]{}

\newcommand{\neqa}{\nonumber\end{eqnarray}}
\newcommand{\la}[1]{\label{#1}}

\newcommand{\eq}[1]{(\ref{#1})}

\def\tr{{\rm tr~}}

\renewcommand{\d}{\partial}

\newcommand{\<}{{\langle}}
\renewcommand{\>}{{\rangle}}

\newcommand{\re}{\relax{\rm I\kern-.18em R}}

\def\su2{{SU(2)}}

\def\[{\left[}
\def\]{\right]}

\def\l{\lambda}

\def\s{\sigma}

\def\[{\left[}
\def\]{\right]}

\def\<{\langle}
\def\>{\rangle}

\def\i2{\frac{i}{2}}

\usepackage{varioref}
\usepackage{makeidx}
\makeindex

\usepackage[english]{babel}
\begin{document}


\title{Integrability for  the  Full Spectrum of Planar AdS/CFT}

\author{ Nikolay Gromov}
\affiliation{DESY Theory, Hamburg, Germany \& II. Institut f\"{u}r Theoretische Physik Universit\"{a}t, Hamburg, Germany \&\\ St.Petersburg INP, St.Petersburg, Russia }
\author{Vladimir Kazakov}
\affiliation{Ecole Normale Superieure, LPT,  75231 Paris CEDEX-5, France $\&$ l'Universit\'e Paris-VI, Paris, France; }
\author{Pedro Vieira}
\affiliation{Max-Planck-Institut f\"ur Gravitationphysik, Albert-Einstein-Institut,  14476 Potsdam, Germany \& \\Centro de F\'\i sica do Porto,  Faculdade de Ci\^encias da Universidade do Porto, 4169-007 Porto, Portugal }

\begin{abstract}
We present a set of functional equations defining the anomalous dimensions of arbitrary local single trace operators  in planar \({\cal N}=4\) SYM theory. It takes the form of a Y-system based on the integrability of the dual superstring \(\s\)-model on the \(AdS_5\times S^5\) background. This Y-system passes some very important tests: it incorporates the full asymptotic  Bethe ansatz at large length of operator \(L  \), including the dressing factor, and it confirms all  recently found wrapping corrections.
The recently proposed $AdS_4/CFT_3$ duality is also treated in a similar fashion.
 \end{abstract}

\maketitle

\section{Introduction}
In the last few years, there has been an impressive progress in computing the spectrum of  anomalous dimensions of planar  \({\cal N}=4\) supersymmetric Yang-Mills (SYM) theory. A great deal of this success was based on Maldacena's AdS/CFT correspondence between this 4D theory and type IIB superstring theory on the \(AdS_5\times S^5\) background \cite{Maldacena:1997re}, and on the integrability  discovered  and exploited on both sides of the correspondence \cite{Minahan:2002ve,Beisert:2003tq,Bena:2003wd,Kazakov:2004qf,Arutyunov:2004vx,Staudacher:2004tk,Beisert:2005tm,Janik:2006dc,Hernandez:2006tk}.
As an outcome, a system of asymptotic Bethe ansatz (ABA) equations was formulated in \cite{Beisert:2006ez} which made   possible the computation of anomalous dimensions of
single trace operators consisting of an asymptotically large number of elementary fields of \({\cal N}=4\) SYM, at any value of the 'tHooft coupling \(\l \equiv 16 \pi ^2 g^2\). This is a very important, though still  limited, information on the non-perturbative behaviour of the  theory.

A far richer and instructive set of quantities to evaluate would be  the anomalous dimensions of ``short" operators such as the famous Konishi operator. The Thermodynamic Bethe ansatz (TBA) approach to the superstring sigma model \cite{Ambjorn:2005wa} has lead to a remarkable calculation of wrapping effects at weak coupling. The 4-loop anomalous dimension of Konishi and similar  operators have been calculated \cite{Bajnok:2008bm}, in complete agreement with the direct perturbative computations \cite{Fiamberti:2008sh}.

Here we propose a  set of equations, the so called Y-system \cite{Zamolodchikov:1991et}, defining the anomalous dimensions of {\it any} physical operator of planar \({\cal N}=4\) SYM at \textit{any} coupling \(g\).
Its integrability properties are those of the discrete classical Hirota dynamics.

The derivation of this Y-system from the bound states of the ABA will be given in a future publication \cite{GKKV:2009}. Here we will demonstrate the crucial test of its selfconsistency: we will see that the Y-system incorporates the ABA equations of \cite{Beisert:2006ez}, including the crossing relation constraining the dressing factor \(S_0\) of the factorized scattering.
We also reproduce the L\"uscher formulae recently used to compute the SYM leading wrapping corrections. In particular we re-derive all known wrapping corrections for twist two operators at weak coupling and present an explicit formula for such corrections for a generic single trace operator of planar $\mathcal{N}=4$.  In the last section we apply our method to the study of the recently conjectured $AdS_4/CFT_3$ duality \cite{ABJM} and find there a new wrapping correction.

Our Y-systems opens  a way to the systematic study of anomalous dimensions of all operators. An even better formulation would be a DdV-like integral equation, in the spirit of the one found in \cite{Gromov:2008gj} for the \(O(4)\) sigma model. This problem is currently under investigation.
\begin{figure}
\includegraphics[width=70mm]{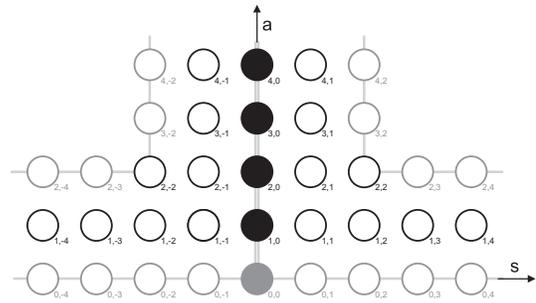}
\caption{\textbf{T}-shaped ``fat hook"  for Y- and T-systems \cite{footnote}. The  middle double line separates the two subgroups with extended \(SU(2|2)_{L}\) and \(SU(2|2)_{R} \) symmetries.
}\label{FatHook}
\end{figure}

\section{Y-system for AdS/CFT}
We will now propose the Y-system which yields the exact planar spectrum of $AdS/CFT$. The Y-system
is a set of functional equations for  functions \(Y_{a,s}(u)\) of the spectral parameter
\(u\) whose indices take values on the lattice represented in Fig.\ref{FatHook}. 
The equations take the usual universal form
\begin{equation}
\label{eq:Ysystem} \frac{Y_{a,s}^+ Y_{a,s}^-}{Y_{a+1,s}Y_{a-1,s}} =\frac{(1+Y_{a,s+1})(1+Y_{a,s-1})}{(1+Y_{a+1,s})(1+Y_{a-1,s})} \,.
\eeq
Throughout the paper we denote $f^{\pm}=f(u\pm i/2)$ and  $f^{[a]}=f(u+ia/2)$.
At the boundaries of the fat-hook we have $Y_{0,s}=\infty$, $Y_{2,|s|>2}=\infty$ and $Y_{a>2,\pm 2}=0$. The product \(Y_{23}Y_{32} \) should be finite so that \(Y_{2,\pm 2}\) are finite.

The anomalous dimension of a particular operator (or the energy of  a string state  in the AdS context) is defined through the corresponding solution of the Y-system and is given by the formula ($E=\Delta-J$)
\beq\la{energy}
\!E=\sum_{j}\epsilon_1(u_{4,j})+\sum_{a=1}^\infty\int_{-\infty}^{\infty}\frac{du}{2\pi i}\,\,\frac{\d\epsilon_a^*}{\d u}\log\left(1+Y_{a,0}^*(u)\right) .\!
\eeq
In terms of \(x(u)\) defined by \(u/g=x+1/x\), the energy dispersion relation reads \(\epsilon_{a}(u)=a+\frac{2ig}{x^{[+a]}}-\frac{2ig}{x^{[-a]}} \), evaluated in the physical kinematics i.e.
for $|x^{[\pm a]}|>1$, while \(\epsilon_a^*(u)\) is given by the same expression evaluated in the mirror kinematics where $|x^{[s]}|>1$ for $a\ge s\ge -a+1$ and $|x^{[-a]}|<1$ \cite{Bajnok:2008bm}. Similarly the asterisk in $Y_{a,0}^*$ indicates that this function should also be evaluated in mirror kinematics. Finally, the Bethe roots are defined by the finite \(L\) Bethe equations
\beq\label{eq:CBE}
Y_{1,0}(u_{4,j})=-1\;,
\eeq
where this expression is evaluated at physical kinematics.

The Y-system is equivalent to an  integrable discrete dynamics on a T-shaped ``fat hook"
drawn in Fig.\ref{FatHook}
given by Hirota equation \cite{footnote}
 \begin{eqnarray}
 \label{Tsystem}
&&\displaystyle T_{a,s}^+T_{a,s}^- =T_{a+1,s}T_{a-1,s}+T_{a,s+1}T_{a,s-1} \,, \\
\label{eq:Ydef}
&&\displaystyle  \text{where} \,\,\, Y_{a,s}=\frac{T_{a,s+1}T_{a,s-1}}{T_{a+1,s}T_{a-1,s}}\;\; .
\end{eqnarray}
The non-zero \(T_{a,s}\)
are represented by all visible circles in Fig.\ref{FatHook}. Hirota equation is invariant w.r.t.
the gauge transformations \(T_{a,s}\to g_1^{[a+s]} g_2^{[a-s]} g_3^{[s-a]} g_4^{[-a-s]} T_{a,s}\). Choosing an appropriate gauge we can impose \(T_{0,s}=1\).


Both the $Y$ and the $T$ systems are infinite sets of functional equations which must still be supplied by certain boundary conditions and analyticity properties. Alternatively, we can identify the proper large $L$ solutions to these equations and find  $T$ and $Y$ functions at finite \(L\) by continuously deforming from this limit  \cite{Gromov:2008gj}. Hopefully  this deformation is unique, as in   \cite{Gromov:2008gj}. Such a numerics can be done by means of an integral DdV-like equation or by some sort of truncation of the Y-system equations.  

\section{Large $L$ solutions and ABA}

We expect the Y-functions  to be smooth and regular  at large \(u\): \( Y_{a,s\ne 0}(u\to \infty)\to const    \), whereas for the black, momentum carrying nodes in Fig.\ref{FatHook},   we impose the asymptotics 
\beq
Y^*_{a\ge 1, 0} \sim \left(\frac{x^{[-a]}}{x^{[+a]}}\right)^{L}   \la{asum}
\eeq
for large $L$ or $u$. As we will now show these asymptotics are consistent with the Y-system \eqref{eq:Ysystem}.
  Indeed, when $L$ is large \(Y_{ a, 0}\,\) goes to zero and we can drop the denominator in the r.h.s. of \eq{eq:Ysystem} at \(s=0\).
  Using \(1+Y_{a,s}=\frac{T_{a,s}^+T_{a,s}^-}{T_{a+1,s}T_{a-1,s}}\) following
from \eqref{Tsystem}-\eqref{eq:Ydef}, we have \beq
\frac{Y_{a,0}^+Y_{a,0}^-}{Y_{a-1,0}Y_{a+1,0}}\simeq\left(\frac{T_{a,1}^+T_{a,1}^-}{T_{a-1,1}T_{a+1,1}}\right)\left(\frac{T_{a,-1}^+T_{a,-1}^-}{T_{a-1,-1}T_{a+1,-1}}\right)\,,
\eeq
where in the equation for $a=1$ one should replace $Y_{0,0}$ by $1$ as can be seen from \eq{eq:Ysystem}.
From
our study of the $O(4)$ $\sigma$-model \cite{Gromov:2008gj} we expect
that $T_{a,s\leq 0}$ and $T_{a,s\geq 0}$ cannot be simultaneously finite
as $L\to\infty$. However, in this limit the full T-system splits into two
independent \(SU(2|2)_{R,L}\) subsystems and, noticing that each factor in the r.h.s. is gauge invariant,
we can always choose  finite solutions
 $T^R_{a,s\leq 0}$  and $T^L_{a,s\geq 0}$
and interpret them as one solution of the full T-system in two different
gauges  (see \cite{Gromov:2008gj} for more details). These are the transfer matrices associated to the rectangular representations of $SU(2|2)_{R,L}$, described in detail in the next section and in the appendix.

The general solution of this discrete 2D Poisson equation in $z$ and $a$ is then
\beq\label{eq:AssY}
Y_{a,0}(u)\simeq\left(\frac{x^{[-a]}}{x^{[+a]}}\right)^{L}
\frac{\phi^{[-a]}}{{\phi^{[+a]}}}\,\,
T^L_{a,-1}T^R_{a,1}
\eeq
where the first two factors
in the r.h.s. represent a zero mode of the discrete Laplace equation
\(\frac{{\cal A}_a^+{\cal A}_a^-}{{\cal A}_{a-1}{\cal A}_{a+1}}=1 \,.
\)
Thus we obtained all $Y_{a,0}$, describing for $a>1$ the AdS/CFT bound states \cite{Dorey:2006dq},  in terms of $T_{a,s}^{L,R}$ up to a
single, yet to be fixed, function $\phi$.  We pulled out the first factor in \eqref{eq:AssY} from the zero mode  to explicitly match
the asymptotics \eq{asum}. The second factor will become the product of fused AdS/CFT dressing factors \cite{Arutyunov:2004vx,Janik:2006dc,Beisert:2006ez} as we shall see below.

%
%
%
\section{Asymptotic transfer matrices}
In the large $L$ limit $Y_{a,0}$ are small and the
whole $Y$-system splits into two  \(SU(2|2)_{L,R}\)  fat hooks on Fig.\ref{FatHook}.
The Hirota equation \eq{Tsystem} also splits into two independent
subsystems. For each of these subsystems there already  exists a  solution
 compatible with the group theoretical interpretation of Y and T-systems: \(T^L_{a,-1}\left(T^L_{1,-s}\right)\) and  \(T^R_{a,1}\left(T^R_{1,s}\right)\) are the transfer matrix eigenvalues of anti-symmetric (symmetric) irreps of the  \(SU(2|2)_L\) and \(SU(2|2)_R\) subgroups of the full $PSU(2,2|4)$ symmetry. It is  known \cite{Tsuboi:1997iq,Kazakov:2007fy}
that these transfer-matrices can be easily generated by the usual fusion procedure.
Explicit expressions for $T_{a,s}$ are given in the Appendix. E.g.,
\begin{eqnarray}
&\!\!\displaystyle\!\!T_{\!1,1}\!\!=\!\!\frac{R^{-\!(\!+\!)} }{R^{-\!(\!-\!)}}\!\!\left[\frac{Q_2^{\!--} \!Q_3^+}{Q_2Q_3^-}\!-\!\frac{R^{-\!(\!-\!)} Q_3^+}{R^{-\!(\!+\!)} Q_3^-}  \!+\!\frac{Q_2^{++} Q_1^-}{Q_2 Q_1^+}\!-\!\frac{B^{+\!(\!+\!)} Q_1^-}{B^{+\!(\!-\!)} Q_1^+} \right] \la{T11}
\end{eqnarray}
where
\(Q_l(u)=\prod_{j=1}^{J_l}(u-u_{l,j})=-R_{l}(u)B_{l}(u) \) and
\beq
\nn  R_l^{(\pm)}(u)\equiv \prod_{j=1}^{K_l} \frac{x(u)-x_{l,j}^{\mp}}{(x_{l,j}^\mp)^{1/2}}\;,\;  B_l^{(\pm)}(u)\equiv\prod_{j=1}^{K_l} \frac{\frac{1}{x(u)}-x_{l,j}^\mp}{(x_{l,j}^\mp)^{1/2}} \,.
\eeq
The index \(l=1,2,3\) corresponds to the roots \(x_{1,j},x_{2,j},x_{3,j}\) (\(x_{7,j},x_{6,j},x_{5,j}\))
for  \(T^L_{\!1,1}\) (\(T^R_{\!1,1}\)) in the  notations of \cite{Staudacher:2004tk}. \(R^{(\pm)}\) and \(B^{(\pm)}\) with no subscript $l$ correspond to the roots \(x_{4,j}\) of the middle node and \(R_{l},B_l\) without supercript $(+)$ or $(-)$ are defined with $x^\pm_j$ replaced by $x_j$. The choice \eqref{T11} is dictated by the condition that   the asymptotic BAE's ought to be reproduced
from the analyticity of  \(T_{1,1}\) at the zeroes \(u_{1,j},u_{2,j},u_{3,j}\) of the  denominators. For \(Q\)-functions of the left and right wings the ABA's read:
\begin{eqnarray}
\left.1\!=\! \frac{Q^+_2B^{(-)}}{Q_2^-B^{(+)}}\right|_{u_{1,k}} \!\!\!\!,\!  \label{BAE}
\left.-1\!=\!\frac{Q_2^{--}Q_1^+Q_3^+}{Q_2^{++}Q_1^-Q_3^-}\right|_{u_{2,k}}\!\!\!\!,\!
\left.1\!=\! \frac{Q_2^+ R^{(-)}}{Q_2^- R^{(+)}}\right|_{u_{3,k}}\!\!\!\!\!.
\end{eqnarray}Once the unknown function $\phi$ is fixed to be
\beq
\frac{\phi^-}{\phi^+}=S^2
\frac{ B^{+(+)}\Rmm}{B^{-(-)}\Rpp}\frac{B_{1L}^{+}
 B_{3L}^{-}}{B_{1L}^{-}  B_{3L}^{+}}\frac{B_{1R}^{+}
 B_{3R}^{-}}{B_{1R}^{-}  B_{3R}^{+}}\label{phi}
\eeq
the Bethe equation \eqref{eq:CBE}  yields the  middle node equation for the full AdS/CFT ABA of \cite{Staudacher:2004tk} at $u=u_{4,k}$\beq
\la{middle}
-1\!=\!\left(\!\frac{x^-}{x^+}\!\right)^{\!\!
L}\!\!\!\left(\!\frac{Q_4^{++}}{Q_4^{--}}\frac{B_{1L}^{-}  R_{3L}^{-}}{B_{1L}^{+}
 R_{3L}^{+}}\frac{B_{1R}^{-}  R_{3R}^{-}}{B_{1R}^{+}
 R_{3R}^{+}}\!\right)^{\!\!\eta}\!\!
\left(\!\frac{ B^{+(+)}}{B^{-(-)}}\!\right)^{\!\!1-\eta} \!\!\!\! S^2,\!\!\!
\eeq
where $\eta=-1$ in the present case and the dressing factor is
$ S(u)=\prod_j\sigma(x(u),x_{4,j})$.
The subscripts $L,R$ refer to the wings.
We will see in the next section that with  the factor \eqref{phi}  $Y_{a0}$
exhibits   crossing invariance and that
this choice of the factor allows to reproduce all  known results for the first wrapping correction of various operators.

\section{scalar factor from crossing }

We will nowsee that the \(Y\)-system constrains the dressing factor by the crossing invariance condition
of \cite{Janik:2006dc}. The S-matrix \(\hat
S(1,2)\) of Beisert
\cite{Beisert:2005tm} admits Janik's crossing relation
which relates the S-matrix with one  argument replaced by
 \(x^\pm\to1/x^\pm\) (particle\(\to\)anti-particle)  to the initial one.
Since the transfer matrices can be constructed as a trace of the product
of S-matrices we expect $Y_{a,0}$ to respect this symmetry.
Indeed, we notice that under the transformation $x^\pm\to1/x^\pm$ (denoted
by $\star$) and
complex conjugation,  $T_{1,1}$ transforms  as $
\overline {T^{\star}_{1,1}}=\frac{Q_1^+Q_3^-}{Q_1^-Q_3^+}\Psi T_{1,1}$, where
$\Psi\equiv \frac{\Rmm\Bpm}{\Rmp\Bpp}$. By demanding the combination $S T_{1,1} \frac{B_{1}^{+}
 B_{3}^{-}}{B_{1}^{-}  B_{3}^{+}}$ to be invariant under that
transformation we  get $\overline{ S^{\star}}=\frac{S}{\Psi}$.
This renders,  using $\frac{\Rmp}{\Bpm}=\frac{\Rpm}{\Bmp} $, the relation $S S^{\star}=\frac{R^{-(+)}B^{-(-)}}{R^{+(+)}B^{+(-)}}$
which is in fact nothing but the crossing
relation for the scalar factor \cite{Janik:2006dc}
\beq
\sigma_{1 2}\sigma_{\bar 12}=\frac{x_2^-}{x_2^+}\frac{x_1^--x_2^-}{x_1^+-x_2^-}\frac{1/x_1^--x_2^+}{1/x_1^+-x_2^+}\;.
\eeq

Note that  crossing does not simply mean $x^\pm \to 1/x^{\pm}$,
but it is also accompanied by an analytical continuation, so one should be
careful with the way the continuation is done because the dressing factor
is a multi-valued function of $(x_1^\pm,x_2^\pm)$.
 Thus
we see that the invariance of
$Y_{1,0}$ imposes  the crossing transformation rule of the dressing factor.
The same invariance property holds for all $Y_{a,0}$.

We conclude that Janik's crossing relation fits nicely
with our Y-system.
The dressing factor is encoded in the Y-system, as for relativistic models (see   \cite{Gromov:2008gj}).

%
%
%

\section{Weak Coupling wrapping Corrections}

Here we will reproduce from our Y-system the results  of \cite{Bajnok:2008bm,Fiamberti:2008sh}
in a rather efficient way and explain how to generalize them to any operator of $\mathcal{N}=4$ SYM. Notice that the large $L$ solution is now completely fixed by \eqref{eq:AssY},\eqref{phi} with the transfer matrices for each $SU(2|2)$ wing generated from $\mathcal{W}$ as explained in the Appendix.

To compute the leading wrapping corrections associated to \textit{any} single trace operator it suffices to plug  the Bethe roots  obtained from the ABA into $Y_{a,0}$ \cite{explanation}. Next we expand this expression for $g\to 0$ and substitute it into the sum \eq{energy}. This ought to be contrasted with the computations in \cite{Bajnok:2008bm},\cite{Bajnok:2008qj} which relied on the explicit form of the S-matrix elements and which are therefore very hard to generalize to generic states.

For example, for the case of two roots $u_{4,1}=-u_{4,2}$ and $L=2$, satisfying the $SL(2)$ ABA ($u_{4,1}=\frac{1}{2\sqrt{3}}+\mathcal{O}(g^2)$), we find
\beq
Y_{a,0}^*=g^8 \left(3\; 2^7\;\frac{3a^3+12au^2-4a}{(a^2+4u^2)^2}\right)^2\frac{1}{y_a(u)y_{-a}(u)}
\eeq
where $y_a(u)=9 a^4-36 a^3+72 u^2 a^2+60 a^2-144 u^2 a-48 a+144 u^4+48 u^2+16$.
Plugging this expression into \eq{energy} we obtain   $(324+864\zeta_3-1440\zeta_5) g^8$, coinciding with the wrapping correction to the anomalous dimension of  Konishi operator $\tr\!\!(ZD^2Z-DZDZ) $ of \cite{Bajnok:2008bm,Fiamberti:2008sh}.

The Konishi state could also be represented as the operator $\tr\!\! \[Z,X\]^2$ in $SU(2)$ sector. To get the ABA equations for the  $SU(2)$ grading we make the following
replacement $T^{su(2)}_{a,s}=
\overline T^{sl(2)}_{s,a}$. The scalar factor \eq{phi} becomes
$\frac{\phi^-}{\phi^+}=S^{2}\frac{Q_4^{++}}{Q_4^{--}}\frac{B_{1L}^{-}  B_{3L}^{+}}{B_{1L}^{+}
 B_{3L}^{-}}\frac{B_{1R}^{-}  B_{3R}^{+}}{B_{1R}^{+}
 B_{3R}^{-}}$
as we can see by matching
with the ABA equations \eq{middle} for $\eta=1$. Repeating the same computation for two magnons, now with $L=4$, we find precisely the same result for wrapping
correction. This is yet another important consistency check of our Y-system.

Another important set of operators are the so called twist two operators for which $L=2$ (in the $SL(2)$ grading) and the Bethe roots are in a symmetric configuration, $u_{4,2j-1}=-u_{4,2j}$ with $j=1,\dots,M/2$. Plugging such configuration into the transfer matrices in the appendix and constructing the corresponding $Y_{a,0}$ from \eqref{eq:AssY} we find a perfect match with the results of \cite{Bajnok:2008qj}.
\begin{figure}
\includegraphics[width=40mm]{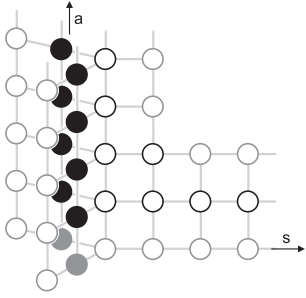}
\caption{``Fat hook"  for $AdS_4/CFT_3$. The $OSp(2,2|6)$ symmetry of the ABJM theory, with two momentum carrying nodes, and the $SU(2|2)$ subgroup is manifest in the diagram.}\label{FatHook2}
\end{figure}

\section{$AdS_4/CFT_3$ correspondence}
The recently conjectured \cite{ABJM} $AdS_4/CFT_3$ correspondence with the ABA formulated in \cite{GVall}, following \cite{MZ2,GVcurve}, can be treated similarly to the $AdS_5/CFT_4$ case. The corresponding Y-system is represented in Fig.\ref{FatHook2}. There are now two sequences of momentum carrying bound-states and the corresponding $Y$-functions are denoted by $Y_{a,0}^4$ and $Y_{a,0}^{\bar 4}$.  At large $L$ we find $Y_{a,0}^4\!\simeq\! \left(\frac{x^{[-a]}}{x^{[+a]}}\right)^{L}
\frac{\phi^{[-a]}_4}{\phi^{[+a]}_4}
T_{a,1}^{su(2)}$, $ Y_{a,0}^{\bar 4}\!\simeq\! \left(\frac{x^{[-a]}}{x^{[+a]}}\right)^{L}
\frac{\phi_{\bar 4}^{[-a]}}{{\phi^{[+a]}_{\bar 4}}}
T_{a,1}^{su(2)}$ where
$\frac{\phi_4^-}{\phi_4^+}=-S_4S_{\bar 4}\frac{Q_4^{++}}{Q_4^{--}}\frac{B_{1}^{-}  B_{3}^{+}}{B_{1}^{+}
 B_{3}^{-}}$ and $\phi_{\bar 4}$ is given by the same expression with $Q_{4}  \to Q_{\bar 4}$. $T_{a,1}$ can be found from the generating functional $\mathcal{W}$ in the appendix replacing $R^{(+)}\!\to\! R^{(+)}_4 R^{(+)}_{\bar 4}$ etc. Finally  \(\epsilon_{a}(u)=\frac{a}{2}+\frac{ih}{x^{[+a]}}-\frac{ih}{x^{[-a]}} \), and in all formulae we should replace $g$ by the interpolating function $h(\lambda)=\lambda+O(\lambda^2)$.
The energy is then computed from an expression analogous to (\ref{energy}) which to leading order at small $\lambda$ yields
\beq
\nonumber E=\sum_{j}\epsilon_1(u_{4,j})+\sum_{j}\epsilon_1(u_{\bar4,j})-\sum_{a=1}^\infty\int_{-\infty}^{\infty}\frac{du}{2\pi} \left(Y_{a,0}^{4*}+ Y_{a,0}^{\bar 4*}\right)
\eeq
Thus, as before we can very easily compute the leading wrapping corrections to any operator of the theory. E.g., for the simplest unprotected length four operator ($L=2$) (irrep ${\bf 20}$, see \cite{MZ2} for details) we find $E=8 h^2(\lambda)-32 \lambda^4+E_{\rm wrapping}\lambda^4+\mathcal{O}(\lambda^6)$ where $E_{\rm
wrapping}=32-16\zeta(2)$.

\section{Appendix: Transfer Matrices}
The $SU(2|2)$ transfer matrices for symmetric ($T_{1,s}$) and antisymmetric ($T_{a,1}$) representations can be found from the expansion of the generating functional \cite{Tsuboi:1997iq,Kazakov:2007fy}
\begin{eqnarray}\nonumber
&&\!\!\!\!\!\!\!\!\!\!\displaystyle{\cal W}=\!\!
\left[1\!-\!\frac{Q_1^{-} B^{+(+)}R^{-(+)} }{Q_1^+B^{+(-)}R^{-(-)} } D\right]\!\!\left[1\!-\!\frac{Q_1^{-} Q_2^{++}R^{-(+)} }{Q_1^+Q_2 R^{-(-)} } D\right]^{-1}\times\\
&&\!\!\!\displaystyle\times\left[1\!-\!\frac{Q_2^{--} Q_3^+R^{-(+)} }{Q_2Q_3^-R^{-(-)} } D\right]^{-1}\!\!\left[1\!-\!\frac{Q_3^+ }{Q_3^-  } D\right]\,\, , \,\, D=e^{-i\partial_u}
\nonumber\\
&&\!\!\!\!\!\!\!\text{as}\,\,\,
{\cal W}=\sum_{s=0}^\infty
T_{1,s}^{[1-s]}D^s \,\, , \,\,
{\cal W}^{-1}=\sum_{a=0}^\infty (-1)^a
T_{a,1}^{[1-a]}D^a\;.
\end{eqnarray}
It can be checked that the transfer matrices $T_{a,1}$ are functions of $x^{[\pm a]}$ alone ($T_{1,s}$ depend on all $x^{[b]}$, $b=-a,-a+2,\dots,a$  ). The transfer matrices for other
representations can be obtained from these  by use of the Bazhanov-Reshitikhin formula \cite{Bazhanov:1989yk}. \\

{\bf Acknowledgments}

The work of NG was partly supported by the German Science Foundation (DFG) under
the Collaborative Research Center (SFB) 676 and RFFI project grant 06-02-16786.  The work  of VK was partly supported by  the ANR grants INT-AdS/CFT (ANR36ADSCSTZ)  and   GranMA (BLAN-08-1-313695) and the grant RFFI 08-02-00287. PV would like to that SLAC for hospitality during the concluding period of writing this paper. We  thank R.Janik and A.Kozak for discussions.

\end{document}